\documentclass[aps,preprint]{revtex4}
\usepackage{amsfonts}
\usepackage{amssymb}
\usepackage{amsmath}
\usepackage{graphicx}
\usepackage{hyperref}
\usepackage{float}
\usepackage{placeins}
\usepackage[font={footnotesize,it}]{caption}
\usepackage{latexsym}
\usepackage{xcolor}
\usepackage{hyperref}

\usepackage{bm}
\begin{document}

\title{Directional Quantum Singularities in Curzon Spacetime}

\author{Mert Mangut}
\email{mert.mangut@emu.edu.tr}
\affiliation{AS237, Department of Physics, Eastern Mediterranean
University, 99628, Famagusta, North Cyprus via Mersin 10, Turkey}
\author{\"{O}zay G\"{u}rtu\u{g}}
\email{ozaygurtug@beykoz.edu.tr}
\affiliation{Beykoz University, Faculty of Engineering and Architecture,
34810, Istanbul -Turkey}
\author{Mustafa Halilsoy}
\email{mustafa.halilsoy@emu.edu.tr}
\affiliation{Department of Physics, Eastern Mediterranean
University, 99628, Famagusta, North Cyprus via Mersin 10, Turkey}

\begin{abstract}
The scalar quantum probe method developed by Horowitz and Marolf  is applied to the cylindrically symmetric Curzon solution. The main cause for choosing the Curzon solution is that it is the best known example that exhibits directional singularity. Interestingly the singularity at $r=0$, for the uncharged Curzon spacetime, which is classically  very strong  with a divergence rate of the order $\frac{1}{r^{10}}$ becomes regular when examined using scalar quantum field. The charged Curzon spacetime, however, due to the emergence of a second singularity off the $r=0$ singularity does not regularize quantum mechanically. All three different charged versions, i.e. electric, magnetic and dyonic share the same feature.

\end{abstract}

\maketitle

\section{Introduction}

The issue of spacetime singularities in Einstein's theory of relativity remains one of the most active areas of research, particularly in relation to the question of the theory's predictability. The key point here is that if singularities are naked, general relativity cannot predict what happens beyond them---the theory effectively breaks down. To address this problem, Roger Penrose proposed the Cosmic Censorship Conjecture, which suggests that all singularities are concealed behind event horizons, thereby preserving determinism for external observers.
However, this conjecture remains unproven, and ongoing research in quantum gravity \cite{1,2,3,4,5,6,7,8}, continues to seek a fundamental understanding or resolution of spacetime singularities. \\

The singularity problem also arises within Weyl's class of metrics, which represent static and axially symmetric solutions to Einstein's field equations. These spacetime are time-independent and possess rotational symmetry around a fixed axis, and they are typically expressed in the Weyl metric form, providing a general framework for obtaining all vacuum solutions with such symmetries. Within this class, two notable examples are the Chazy-Curzon \cite{9,10}, which we shall refer in this study as the Curzon solution and Levi-Civita \cite{11} solution. The Curzon solution represents the gravitational field of a point-like mass in an axially symmetric, static spacetime; unlike the Schwarzschild solution, it lacks an event horizon and instead exhibits a naked singularity along the symmetry axis, emphasizing the limitations of spherical symmetry. The Levi-Civita solution, on the other hand, describes the field of an infinite static line mass with cylindrical symmetry. Its singularity structure depends on a parameter related to the linear mass density: for most values, the axis $\rho = 0$ represents a curvature singularity, while in certain limits it corresponds to a cosmic string--like conical singularity. Together, these solutions illustrate how different symmetry assumptions in static vacuum spacetimes lead to distinct geometric and singularity structures within general relativity. \\

In this study, we aim to investigate the formation and properties of timelike naked directional singularities from a quantum mechanical perspective in both charged and uncharged Curzon spacetimes.
 Classical singularity analysis and geodesic motion of the Curzon spacetime, which is not our aim here, had been investigated long ago by Scott and  Szekeres \cite{H1,H2}. For a detailed review of cylindrically symmetric spacetime we refer the readers to Bronnikov et. al. \cite{H3}. \\

In cylindrical coordinates $(t, \rho, z, \varphi)$ the Curzon solution is described by the line element\\

\begin{equation}
ds^{2}=-e^{\frac{-2 m}{\sqrt{\rho^{2}+z^{2}}}} d t^{2}+e^{\frac{2 m}{\sqrt{\rho^{2}+z^{2}}}} \left[e^{\frac{-m^2\rho^2}{(\rho^{2}+z^{2})^2}}\left(d \rho^{2}+d z^{2}\right)+\rho^{2} d \varphi^{2}\right],
\end{equation}
in which $m$ stands for the mass parameter. Throughout the paper we shall use the unit system in which $G=$ Newton's constant $=1$ and $c=$ speed of light $=1$, so that the Einstein's equations will take the form $G_{\mu\nu}=8\pi T_{\mu\nu}$. More appropriately, for our present purpose we transform this line element to the spherical coordinates through $\rho=r \sin \theta, z=r \cos \theta$, to obtain

\begin{equation}
ds^{2}=-e^{-\frac{2 m}{r}} d t^{2}+e^{\frac{2 m}{r}}\left[e^{-\frac{m^{2} \sin ^{2} \theta}{r^{2}}}\left(d r^{2}+r^{2} d \theta^{2}\right)+r^{2} \sin ^{2} \theta d \varphi^{2}\right].
\end{equation}

It is easily seen that for $m \rightarrow 0$, or $ r \rightarrow \infty$ we get the flat metric in spherical coordinates. The charged version of the Curzon solution can be  described by the following line element 

\begin{equation}
d s^{2}=-\frac{4}{K^{2}} d t^{2}+\frac{K^{2}}{4}\left[e^{-\frac{m^{2} \sin ^{2} \theta}{r^{2}}}\left(d r^{2}+r^{2} d \theta^{2}\right)+r^{2} \sin ^{2} \theta d \varphi^{2}\right],
\end{equation}
where $K=(1+p) e^{m / r}+(1-p) e^{-m / r}$. The new constant $p$ is related to the charge $q$ of the source through

\begin{equation}
p^2=1+q^2
\end{equation}
so that $p \geqslant 1$, as $0 \leqslant q<\infty$. The Curzon limit is obtained simply by setting $p=1$. From the physics point of view the Curzon spacetime can have sources in three different forms; pure electric, pure magnetic or both, i.e. dyonic. To see this we consider an electromagnetic vector potential ansatz

\begin{equation}
A_{\mu}=-\frac{C_{0}}{K}\left(\frac{p-1}{p+1}\right) e^{-m / r} \delta_{\mu}^{t}+C_{1} \cos \theta \delta_{\mu}^{\varphi},
\end{equation}
in which $C_{0}$ and $C_{1}$ are constants to be fixed by the Einstein-Maxwell (EM) equations. First we check that the vacuum Maxwell equations $\nabla_{\mu} F^{\mu\nu}=0=\nabla_{\mu}\overset{\ast}{F}{}^{\mu\nu}$ are satisfied where $\overset{\ast}{F}{}^{\mu\nu}$ stands for the dual field tensor. The EM equations

\begin{equation}
R_{\mu}{ }^{\nu}=8\pi T_{\mu}{ }^{\nu},
\end{equation}
are satisfied for the Ricci tensor $R_{\mu}{ }^{v}$ and the energy-momentum tensor given by

\begin{equation}
T_{\mu}{ }^{\nu}=\frac{16 m^{2} q^{2} e^{\frac{m^2 \sin ^{2} \theta}{r^{2}}}}{r^{4} K^{4}} \operatorname{diag}(-1,-1,1,1)=\operatorname{diag}\left(-\rho, p_{r}, p_{\theta}, p_{\varphi}\right).
\end{equation}
Our notation is such that the energy density $\rho$ and pressure components are given by

\begin{equation}
\begin{aligned}
\rho & =-p_{r}=p_{\theta}=p_{\varphi} \\
& =\frac{2 m^{2} q^{2}}{\pi r^{4} K^{4}} e^{\frac{m^{2} \sin ^{2} \theta}{r^{2}}}.
\end{aligned}
\end{equation}

It can easily be checked that all conditions, i.e. Weak, Null, Strong and Dominant  are satisfied showing that the Curzon spacetime is established on firm physical ground. The EM equations (6) are satisfied, provided

\begin{equation}
(1-p)^{2} C_{0}^{2}+\frac{4}{m^{2}} C_1^{2}=8\left(p^{2}-1\right)
\end{equation}
holds. From this constraint condition for a dyonic solution we obtain the constants $C_{0}$ and $C_{1}$ for pure electric and pure magnetic solutions as $C_{0}=2 \sqrt{2}\left(\frac{1+p}{p-1}\right)^{1 / 2}$ and $C_1=\sqrt{2} m\left(p^{2}-1\right)^{1 / 2}$, respectively. We note that charged Curzon solution was also considered  by different authors \cite{13,14}.\\

Obviously the Curzon spacetime, both charged and uncharged possess a strong singularity at $r=0$, which makes the main subject matter of the present study. The Kretchmann scalar of the solution (3) is given by

\begin{equation}
\mathcal{K}(r,\theta) = 
\frac{16 m^2 e^{\frac{2 m \left(m \sin ^2(\theta )+2 r\right)}{r^2}}}{r^{10} \left((p+1) e^{\frac{2 m}{r}}-p+1\right)^8}\;
\mathcal{P}(r,\theta).
\end{equation}

The regular polynomial $P(r, \theta)$ is given explicitly in Appendix $A$. Setting $p=1$, we obtain the Kretchmann scalar for the vacuum Curzon solution. It has bean known for a long time that the singularity at $r=0$ for the vacuum Curzon solution exhibits directional properties. The singular behavior of the Kretchmann scalar depends on the direction by which the singular point is approached. For the vacuum Curzon solution, if the singularity $r=0$ is approached from  the equatorial plane $\theta=\pi / 2$, the Kretchmann scalar diverges, indicating true curvature singularity. However, if the singularity  is approached along the $z$-axis, namely $\theta=0$ or $\theta=\pi$, the Kretchmann scalar remains  regular \cite{12}.\\

In the case of the charged Curzon solution, the Kretchmann scalar diverges at two different locations. One is at $r=0$ and the other is at $r\equiv r_{\star}=\frac{2 m}{\ln \left(\frac{p-1}{1+p}\right)}$. As in the case of vacumm Curzon solution, these points becomes singular if they are approached along the equatorial plane $(\theta=\pi / 2)$.  Approaching these points along the $z$-axis scalar invariant remains regular and hence possesses directional behaviour.\\

In this study, we investigate the classical singularities present in both the charged and uncharged Curzon solutions from a quantum mechanical perspective. Our approach focuses on the propagation of a scalar field, which plays a central role in quantum field theory on curved spacetime backgrounds. By examining the behavior of the scalar field near these singularities, we aim to gain insight into their nature and explore the possibility of their resolution within a quantum framework. To this end, we adopt the approach developed by Wald \cite{15} for handling dynamics in nonglobally hyperbolic, static spacetimes, which was later extended by Horowitz and Marolf (HM) \cite{16}. This method, which employs quantum wave packets to probe spacetime singularities, has proven particularly effective in assessing the quantum nature of classically singular spacetimes \cite{17,18,19,20,21,22}. In addition to these investigations, the classical naked singularities in Levi-Civita solutions have also been examined \cite{23}. It has been demonstrated that such classical naked singularities can be smoothed out by choosing appropriate parameters within the Levi-Civita solution. Furthermore, the classical naked singularities in the Zipoy--Voorhees (ZV) solution \cite{H4,H5}, which describes the exterior geometry of non-spherical compact objects, have been considered in \cite{24}. The singularity structure of the ZV solution shares a similar directional character with that of the Curzon solution, and it has been shown that this singularity can be healed when probed with a specific mode. \\ 
Accordingly, we employ the HM criterion as the central tool in our analysis. It is worth to empasize here that in the charged Curzon solution, our target will be the outermost singularity located at $r_{\star}=\frac{2 m}{\ln \left(\frac{p-1}{p+1}\right)}$.

Prior to employing the HM method we have to show that the singularity is a time like, naked one. For this purpose, we consider a general singular hypersurface described by $\phi(r)=r-r_{0}$, $\left(r_{0}=\right.$ constant $)$. The square of the normal vector to this surface is $n^{2}=\left(\nabla\phi \right)^2=g^{rr}\left(\phi^{\prime} \right)^2$, $\left(\phi^{\prime}=\frac{d \phi}{d r}\right)$ which is $(+)$ definite as $r \rightarrow 0$. Since the normal vector is spacelike it shows that the singular surface containing  $r=0$ is timelike, apt for applying the HM technique. For the charged Curzon case the same analysis applies for the hypersurface $\phi=r-r_{\star}$ which is also timelike so that we can use the method of HM.

\section{Quantum Singularity Analysis }\label{2}

Within general relativity, the emergence of spacetime singularities represents a breakdown of predictability. The situation is particularly critical when such singular regions are not shielded by horizons, giving rise to naked singularities accessible to distant observers. In these domains, where curvature scalars diverge, classical analyses become inapplicable, and quantum corrections are indispensable. Although a complete framework for quantum gravity is still absent, adopting quantum-inspired approaches to examine singular behavior constitutes a constructive direction.\\

A significant step in this direction was Wald's proposal \cite{15}, which introduced a physically meaningful prescription to probe singularities using a massless Klein--Gordon field in static spacetimes with timelike singularities. In this approach, the classical particle probe is replaced by a quantum scalar wave packet, and the problem reduces to studying the self-adjointness of the spatial wave operator. Building on this idea, Horowitz and Marolf \cite{16} argued that such a spacetime is quantum mechanically nonsingular if the evolution of all states is uniquely determined for all times; otherwise, a failure of uniqueness signals quantum singularity.\\

To formalize the analysis, we examine the dynamics of a wave packet subject to the massless Klein--Gordon equation  $(\nabla^{\mu}\nabla_{\mu})\psi=0$ on a static background lacking global hyperbolicity. By isolating the temporal component, the problem reduces to analyzing the spatial part of the Klein--Gordon operator. The issue of singularity resolution is then translated into the mathematical question of whether this spatial operator admits a unique self-adjoint extension, ensuring well-defined unitary evolution. For this purpose,  the massless Klein--Gordon equation can be written as

\begin{equation}
\frac{\partial ^{2}\psi }{\partial t^{2}}=A\psi,
\end{equation}
in which $A$ represents the spatial operator. Furthermore, the generic operator solution of Eq.(11) is given by

\begin{equation}
\psi \left( t\right) =e^{-it\sqrt{A_{E}}}\psi \left( 0\right),
\end{equation}
where $A_E$ denotes a self-adjoint extension of the operator introduced in Eq.(11). The central question is whether $A$ has a unique extension that guaranties the well-posedness of the associated Cauchy problem. A unique extension ensures unitary and deterministic time evolution, corresponding to quantum regularity. In contrast, non-uniqueness implies ambiguity in the choice of extension, leading to quantum singularity. In this context, the uniqueness of the extension of $A$ is determined within von Neumann's framework of self-adjoint extensions through  analysis of its deficiency spaces. This criterion can be formulated in terms of deficiency indices, expressed as

\begin{equation}
\left( A^{\ast }\pm i\right) \psi =0,
\end{equation}
and the solution of Eq.(13) admits no square-integrable functions, implying that it does not belong to the Hilbert space. Therefore, the operator $A$ has a unique self-adjoint extension, and is thus essentially self-adjoint.

For the charged Curzon spacetime, when the standard massless Klein--Gordon equation is written in the form of Eq.(11), the operator $A$ to be used in the analysis is obtained as

\begin{equation}
A = -\left\{
\frac{16}{K^{4}r e^{2\gamma}}\!\left( 2 \frac{\partial}{\partial r} + r \frac{\partial^{2}}{\partial r^{2}} \right)
+ \frac{16}{K^{4}r^2 \sin^{2}\theta}\,\frac{\partial^{2}}{\partial \varphi^{2}}
+ \frac{16}{K^{2}r^2 e^{2\gamma}}\!\left( \cot\theta \frac{\partial}{\partial \theta} + \frac{\partial^{2}}{\partial \theta^{2}} \right)
\right\}.
\end{equation}

Plugging  Eq.(14) into Eq.(13) and assuming the azimuthal separation ansatz of the form  $\psi=f(r, \theta) e^{ \pm i k \varphi}$  , one finds

\begin{equation}
\left\{\frac{r}{e^{2\gamma}}\left(2 \frac{\partial}{\partial r}+r^2 \frac{\partial^2}{\partial r^2}\right)\mp\frac{k^2}{\sin ^2 \theta}+\frac{1}{e^{2\gamma}}\left(\cot \theta \frac{\partial}{\partial \theta}+\frac{\partial^2}{\partial \theta^2}\right)\pm \frac{K^4 r^2 i}{16}\right\} f(r, \theta)=0.
\end{equation}

Considering the angular dependence of the naked singularity, the angular orientation of the quantum probe plays a critical role. Accordingly, the massless wave must be directed towards the naked singularity along the angular direction $\theta=\pi/2$. By imposing this condition on the general wave function, Eq.(15) becomes

\begin{equation}
\frac{d^2 R(r)}{d r^2}+\frac{2}{r} \frac{d R(r)}{d r}\mp\left[\frac{k^2}{r^2} e^{2 \gamma}-\frac{K^2 e^{2 \gamma} i}{16}\right] R(r)=0,
\end{equation}
in which

\begin{equation}
\begin{aligned}
& K(r, \pi / 2)=(1-p) e^{-\frac{m}{r}}+(1+p) e^{\frac{m}{r}} \\
& 2 \gamma(r, \pi / 2)=-\frac{m^2}{r^2} .
\end{aligned}
\end{equation}

Here, $f(r, \theta=\pi/2)\equiv R(r)$. It is important to note at this stage that the HM criterion requires considering the entire space from singular point to infinity. For this reason, in the following subsections, we provide a detailed analysis of the square-integrability properties of the solutions to Eq.(16) in the asymptotic limits $r\rightarrow\infty$ and $r\rightarrow0$, respectively.

\subsection{In the case of $r\rightarrow\infty$}

In this asymptotic case, the generic metric functions behave as

\begin{equation}
\begin{aligned}
& K(r, \pi / 2)\approx 2\left(1+\frac{pm}{r} \right), \\
& 2 \gamma(r, \pi / 2)\approx 0.
\end{aligned}
\end{equation}

By substituting the generic asymptotic forms from Eq.(18) and keeping only the dominant terms, the radial equation (16) reduces to

\begin{equation}
\frac{d^2 R(r)}{d r^2}+\frac{2}{r} \frac{d R(r)}{d r}\pm\ i R(r)=0.
\end{equation}

The solution of Eq.(19) is given by

\begin{equation}
R(r)=\frac{C_1}{r}sin(\kappa r)+\frac{C_2}{r}cos(\kappa r),
\end{equation}
in which $C_1$ and $C_2$ are integration constants. In this case, $\kappa=\pm i $. The square-integrability property of the solution to Eq.(20) is examined with respect to the general square norm  defined as \cite{17}

\begin{equation}
\Vert R\Vert ^{2}=\int g_{tt}^{-1}\sqrt{-h}RR^{\ast }dr,
\end{equation}
where, $g_{tt}$ denotes the $(-tt)$ component of the space--time metric, $h$ represents the determinant of the spatial part of the line element, and $\ast$ indicates the complex conjugate of the radial solution.\\

Plugging  Eq.(18) and Eq.(20) into Eq.(21), the square norm becomes

\begin{equation}
\Vert R\Vert ^{2}=\int_{const.}^{\infty}\left(1+\frac{pm}{r} \right)^4 \left[\sin \left(\sqrt{2} r\right)+\cosh \left(\sqrt{2} r\right)\right]dr.
\end{equation}

Here, for clarity in establishing the divergent character of the norm, the integration constants are fixed to $C_1=C_2=1$, which entails no loss of generality in the formal framework. Using the comparison test, we now show that the integral obtained in Eq.(22), although rather complicated in appearance, is divergent. For this purpose, the integrand function is estimated from below as follows

\begin{equation}
\frac{(1+pm)^4}{r^4}\left(cosh(\sqrt{2}r)-1\right)\underset{r\rightarrow\infty}{<} \left(1+\frac{pm}{r} \right)^4 \left[\sin \left(\sqrt{2} r\right)+\cosh \left(\sqrt{2} r\right)\right].
\end{equation}

Integrating the lower bounding function yields

\begin{equation}
\begin{aligned}
&\int_{const.}^{\infty}\frac{(1+pm)^4}{r^4}\left(cosh(\sqrt{2}r)-1\right)dr=-\frac{(1+pm)^4}{3r^3} \vert_{const.}^{\infty}\\
&+\frac{(1+pm)^4\left(\sqrt{2} r \left(2 r^2 \text{Shi}\left(\sqrt{2} r\right)-\sinh \left(\sqrt{2} r\right)\right)-2 \left(r^2+1\right) \cosh \left(\sqrt{2} r\right)\right)}{6 r^3}\vert_{const.}^{\infty},
\end{aligned}
\end{equation}
in which, $\text{Shi}$ represents the hyperbolic sine integral, which diverges $r\rightarrow\infty$. Since the integral obtained in Eq.(24) is divergent, the norm integral also diverges by the comparison test. Consequently, the solution in the limit $r\rightarrow\infty$ is not square-integrable $(\Vert R\Vert ^{2}\rightarrow\infty)$ and does not belong to the Hilbert space. 

\subsection{In the case of $r\rightarrow r_{\star}$}

As previously discussed, our attention will be directed toward the outermost singularity situated at $r=r_{\star}$. To render the radial differential equation more amenable to analysis, it is convenient to introduce a new variable $x=r-r_{\star}$, thereby allowing the equation to be reformulated in terms of $x$. In the vicinity of the singularity, $x$ is extremely small $(x<<1)$, allowing to neglect second-order terms in $x$, then the radial differential equation reduces to a form that can be solved analytically. In this context, the generic limiting  functions can be written as

\begin{equation}
\begin{aligned}
& K(x, \pi / 2)\approx \left(1-p \right)\left[1+\frac{m}{r_{\star}^2}x \right]e^{-\frac{m}{r_{\star}}}+\left(1+p \right)\left[1-\frac{m}{r_{\star}^2}x \right]e^{\frac{m}{r_{\star}}}, \\
& 2 \gamma(x, \pi / 2)\approx -\frac{m^2}{r_{\star}^2}+\frac{2m^2}{r_{\star}^3}x.
\end{aligned}
\end{equation}

By substituting Eq.(25) in the asymptotic regime, the radial equation (16) reduces to

\begin{equation}
\frac{d^2 R(x)}{d x^2}+\left\{a-bx \right\}\frac{d R(x)}{d x}+\left\{e+dx \right\}R(x)=0.
\end{equation}

Here, the constant coefficients are given by

\begin{equation}
\begin{aligned}
&a=\frac{2}{r_{\star}}, \\
&b=\frac{2}{r_{\star}^2},\\
&e=\frac{k^2}{r_{\star}^2}-\frac{k^2m^2}{r_{\star}^4}+\left( 1-\frac{m^2}{r_{\star}^2}\right)\left(\pm\frac{i}{16} \right)\left\{e^{-\frac{2m}{r_{\star}}}\left(1-p \right)^2+e^{\frac{2m}{r_{\star}}}\left(1+p \right)^2-2q^2 \right\}, \\
&d=\left( 1-\frac{m^2}{r_{\star}^2}\right)\left(\pm\frac{i}{16} \right)\left\{e^{-\frac{2m}{r_{\star}}}\left(1-p \right)^2+e^{\frac{2m}{r_{\star}}}\left(1+p \right)^2-2q^2 \right\}-\frac{k^2}{r_{\star}^3}+\frac{2k^2m^2}{r_{\star}^5}\\
&\pm \left(\frac{m^2i}{8r_{\star}^3} \right)\left\{e^{-\frac{2m}{r_{\star}}}\left(1-p \right)^2+e^{\frac{2m}{r_{\star}}}\left(1+p \right)^2-2q^2 \right\}.
\end{aligned}
\end{equation}

The general solution of Eq.(26) is

\begin{equation}
\begin{aligned}
R(x)=&C_3 e^{\frac{d x}{b}} H_{\frac{e b^2+a d b+d^2}{b^3}}\left(\frac{\sqrt{b} x}{\sqrt{2}}-\frac{a b+2 d}{\sqrt{2} b^{3/2}}\right)\\
&+C_4e^{\frac{d x}{b}}{}_1F_{1}\left(-\frac{a b d+b^2 e+d^2}{2 b^3};\frac{1}{2};\left(\frac{\sqrt{b} x}{\sqrt{2}}-\frac{a b+2 d}{\sqrt{2} b^{3/2}}\right)^2\right).
\end{aligned}
\end{equation}
where $H$ and ${}_1F_{1}$ represent the Hermite polynomial and the Kummer confluent hypergeometric function, respectively. In the limit $r\rightarrow r_{\star}$, where the new coordinate $x=r-r_{\star}$ becomes very small $(x<<1)$, the general solution given in Eq. (28), which involves Hermite polynomial and Kummer confluent hypergeometric functions, can be significantly simplified. Expanding these functions around $x=0$ and keeping only the leading-order terms, the radial function reduces to a linear form,

\begin{equation}
R(x) \;\approx\; A + Bx,
\end{equation}
where the coefficients $A$ and $B$ are given explicitly as
\begin{equation}
A \;=\; c_1\,H_{\nu}\!\left(\tfrac{ab-2d}{\sqrt{2}\,b^{3/2}}\right) 
+ c_2\,{}_1F_{1}\!\left(-\tfrac{\nu}{2}; \tfrac{1}{2}; \tfrac{(ab-2d)^2}{2b^3}\right),
\end{equation}
and
\begin{align}
B \;=\;& c_1\left[ \left(-a+\tfrac{d}{b}\right)\,H_{\nu}\!\left(\tfrac{ab-2d}{\sqrt{2}\,b^{3/2}}\right) 
+ \nu \sqrt{2b}\; H_{\nu-1}\!\left(\tfrac{ab-2d}{\sqrt{2}\,b^{3/2}}\right) \right] \notag \\[6pt]
&+ c_2\left[ \left(-a+\tfrac{d}{b}\right)\,{}_1F_{1}\!\left(-\tfrac{\nu}{2}; \tfrac{1}{2}; \tfrac{(ab-2d)^2}{2b^3}\right) 
- \tfrac{\nu\,(ab-2d)}{b}\;{}_1F_{1}\!\left(1-\tfrac{\nu}{2}; \tfrac{3}{2}; \tfrac{(ab-2d)^2}{2b^3}\right) \right],
\end{align}
with the parameter
\begin{equation}
\nu \;=\;\frac{-b^{3}+c b^{2}-a d b+d^{2}}{b^{3}},
\end{equation}
in which $H_{\nu}$ denotes the Hermite function of order $\nu$. This representation provides a compact linear approximation of the full solution in the neighborhood of $x=0$, which is particularly useful for analyzing the dominant behavior of the system near the singularity.

Plugging Eq.(29) into Eq.(21), one finds

\begin{equation}
\begin{aligned}
&\Vert R\Vert ^{2}\approx\int_{0}^{const.} \left\{\left(r_{\star}^2-m^2 \right)\vert A \vert^2\left\{e^{-\frac{2m}{r_{\star}}}\left(1-p \right)^2+e^{\frac{2m}{r_{\star}}}\left(1+p \right)^2-2q^2 \right\}^2 \right.\\
&\left.+\left[ \left(r_{\star}^2-m^2 \right)\left\{e^{-\frac{2m}{r_{\star}}}\left(1-p \right)^2+e^{\frac{2m}{r_{\star}}}\left(1+p \right)^2-2q^2 \right\}^2 \right.\right.\\
&\textcolor{blue}{+}\left.\left.\frac{4m\left(r_{\star}^2-m^2 \right)}{r_{\star}^2} \left\{e^{-\frac{4m}{r_{\star}}}\left(1-p \right)^4+e^{\frac{4m}{r_{\star}}}\left(1+p \right)^4-4p^2-2q^2e^{-\frac{2m}{r_{\star}}}\left(1-p \right)^2+2q^2e^{\frac{2m}{r_{\star}}}\left(1+p \right)^2 \right\} \right.\right.\\
&\textcolor{blue}{+}\left.\left.\left(\frac{2m^2}{r_{\star}}+2-\frac{m^2}{r_{\star}^2} \right)\left\{e^{-\frac{2m}{r_{\star}}}\left(1-p \right)^2+e^{\frac{2m}{r_{\star}}}\left(1+p \right)^2-2q^2 \right\}^2 \right] \left(A^{\star}B+B^{\star}A \right)\right\}xdx.
\end{aligned}
\end{equation}

The linear integral given in Eq.(33) is simply integrable and belongs to the Hilbert space. As a consequence, the analysis of the entire space reveals that the spatial operator  $A$ is not essentially self-adjoint, and the outermost singularity remains quantum mechanically singular.

\subsection{Uncharged case ($p=1$)}

In the asymptotic limit $r\rightarrow\infty$, we analyze the uncharged configuration of the spacetime by taking $p=1$ in the metric functions defined in Eq.(18), from which the following metric functions are obtained 
\begin{equation}
\begin{aligned}
& K(r, \pi / 2)\approx 2\left(1+\frac{m}{r} \right), \\
& 2 \gamma(r, \pi / 2)\approx 0.
\end{aligned}
\end{equation}
Under this case, the radial equation reduces to
\begin{equation}
\frac{d^2 R(r)}{d r^2}+\frac{2}{r} \frac{d R(r)}{d r}\pm\ i R(r)=0.
\end{equation}
The solution of Eq.(35) has the same functional form as in the previous case and is given by
\begin{equation}
R(r)=\frac{C_1}{r}sin(\kappa r)+\frac{C_2}{r}cos(\kappa r).
\end{equation}
If we substitute Eq.(36) into Eq.(21), the square norm can be written as  
\begin{equation}
\Vert R\Vert ^{2}=\int_{const.}^{\infty}\left(1+\frac{m}{r} \right)^4 \left[\sin \left(\sqrt{2} r\right)+\cosh \left(\sqrt{2} r\right)\right]dr.
\end{equation}   
Similar to the analysis in the previous section, the comparison test indicates that this integral diverges, and consequently, the solution remains non--square-integrable in this limit.
Nevertheless, within this limiting regime the singularity at $r=0$ remains amenable to probing. We now proceed to examine this singularity in detail. In the limit $r\rightarrow0$, the generic metric functions exhibit the following behavior

\begin{equation}
\begin{aligned}
& K(r, \pi / 2)\approx 2e^{\frac{m}{r}}, \\
& 2 \gamma(r, \pi / 2)\approx -\frac{m^2}{r^2}.
\end{aligned}
\end{equation}

Using Eq.(38) in the limit  $r\rightarrow0$, the radial equation (16) simplifies to

\begin{equation}
\frac{d^2 R(r)}{d r^2}+\frac{2}{r} \frac{d R(r)}{d r}\mp\ \frac{k^2}{r^2}R(r)=0.
\end{equation}

The solution to Eq.(39) can be expressed as

\begin{equation}
R(r)=C_3r^{\alpha_1}+C_4r^{\alpha_2},
\end{equation}
where $C_3$ and $C_4$ represent the integration constants. Also, the associated exponents are

\begin{equation}
\begin{aligned}
& \alpha_1=\frac{1}{2}\left(-1-k\sqrt{\frac{1}{k^2}\pm4} \right), \\
& \alpha_2=\frac{1}{2}\left(-1+k\sqrt{\frac{1}{k^2}\pm4} \right) .
\end{aligned}
\end{equation}

Substituting Eq.(38) and Eq.(40) into Eq.(21), and adopting the choice $C_3=C_4=1$ as in the previous section for simplicity, the square norm takes the form

\begin{equation}
\Vert R\Vert ^{2}=\int_{0}^{const.}2^4e^{\frac{4m}{r}-\frac{m^2}{r^2}}r^{1-k \sqrt{\pm4+\frac{1}{k^2}}} \left(r^{k \sqrt{\pm4+\frac{1}{k^2}}}+1\right)^2dr.
\end{equation}

By applying the comparison test, we demonstrate that the integral in Eq.(42) diverges. To this method, we bound the integrand from below as follows:

\begin{equation}
\begin{aligned}
&2^4\left(\frac{4m}{r}-\frac{m^2}{r^2}\right)r^{1-k \sqrt{\pm4+\frac{1}{k^2}}} \left(r^{k \sqrt{\pm4+\frac{1}{k^2}}}+1\right)^2\underset{r\rightarrow0}{<}\\
&2^4e^{\frac{4m}{r}-\frac{m^2}{r^2}}r^{1-k \sqrt{\pm4+\frac{1}{k^2}}} \left(r^{k \sqrt{\pm4+\frac{1}{k^2}}}+1\right)^2.
\end{aligned}
\end{equation}

Upon integrating the lower bounding function, one obtains

\begin{equation}
\begin{aligned}
&\int_{0}^{const.}2^4\left(\frac{4m}{r}-\frac{m^2}{r^2}\right)r^{1-k \sqrt{\pm4+\frac{1}{k^2}}} \left(r^{k \sqrt{\pm4+\frac{1}{k^2}}}+1\right)^2dr=2^4m\left\{8 r-\frac{4 r^{1-k \sqrt{\pm4+\frac{1}{k^2}}}}{k \sqrt{\pm4+\frac{1}{k^2}}-1}\right.\\
&\left.+\frac{m \left(r^{k \left(-\sqrt{\pm4+\frac{1}{k^2}}\right)}-r^{k \sqrt{\pm4+\frac{1}{k^2}}}-2 k \sqrt{\pm4+\frac{1}{k^2}} \ln (r)\right)}{k \sqrt{\pm4+\frac{1}{k^2}}}+\frac{4 r^{k \sqrt{\pm4+\frac{1}{k^2}}+1}}{k \sqrt{\pm4+\frac{1}{k^2}}+1}\right\}\vert_{0}^{const.},
\end{aligned}
\end{equation}

As Eq.(44) yields a divergent integral, the norm integral necessarily diverges by the comparison test. Consequently, in the case where the solution at $r\rightarrow0$ is not square-integrable $(\Vert R\Vert ^2 \rightarrow \infty)$, implies that the spatial operator for the entire space becomes essential self-adjoint, ensuring that the spacetime is quantum mechanically regular.

\section{Discussion}

In this study, we have examined the classical and quantum nature of the singularities present in the Curzon and charged Curzon spacetimes. The Curzon metric, an exact, static, and axisymmetric vacuum solution to Einstein's field equations, possesses a naked timelike singularity that exhibits a distinct directional dependence. Specifically, the Kretschmann scalar remains finite along the symmetry $(z)$ axis while diverging in other directions. The charged extension of this geometry preserves this anisotropic character but introduces an additional singular surface, further enriching the singular structure of the spacetime. \\
To investigate these singularities from a quantum mechanical standpoint, we employed massless scalar fields obeying the Klein--Gordon equation as quantum probes, following the approach initiated by Wald and later developed by Horowitz and Marolf. Since the Klein--Gordon field represents a bosonic field with zero spin, it serves as an appropriate test field for assessing quantum regularity. The directional nature of the Curzon singularity constrains the propagation of these scalar modes primarily to the equatorial plane, allowing a meaningful analysis of their behavior near the singular region. \\
Our results indicate that the strong classical singularity in the vacuum Curzon spacetime becomes quantum mechanically regular when probed with scalar fields, suggesting that quantum effects can effectively "smooth out" this classical divergence. In contrast, for the charged Curzon spacetime, the emergence of an additional singular surface prevents the attainment of quantum regularity. Consequently, the charged Curzon solution remains quantum mechanically singular, emphasizing that the inclusion of charge worsen rather than mitigates the spacetime's pathological features. \\
Future work may involve extending this analysis to include higher-spin quantum fields, such as Dirac $ (spin- \frac{1}{2}) $ or vector $ (spin-1) $ fields, to determine whether the observed regularization persists across different field types. Additionally, examining non-minimally coupled scalar fields or semiclassical backreaction effects could provide further insight into the role of quantum corrections in smoothing directional singularities. A comparative study with other axisymmetric solutions exhibiting anisotropic singular structures may also deepen our understanding of the interplay between classical geometry and quantum field behavior near singularities. Similar analysis can be carried out to spacetimes having sources other than the electromagnetic field. The case of Yang-Mills will be our next project to investigate whether the Einstein-Yang-Mills metric remains singular or not under a scalar quatum probe.

\section*{Acknowledgments}

We wish to thank S. Habib Mazharimousavi for sharing with us the dyonic source of the Curzon solution.

\section*{Appendix A: Explicit form of regular polynomial $\mathcal{P}(r,\theta)$}

The explicit form of regular polynomial $\mathcal{P}(r,\theta)$ is given by

\begin{equation}
\begin{aligned}
& \mathcal{P}(r,\theta)=4 m^4 \sin^2(2\theta) 
  \left( (p-1)^2 - (p+1)^2 e^{\tfrac{4m}{r}} \right)^2  \\[6pt]
& + \Big( 
   - (p-1)^2 \big( m^2 \cos(2\theta) - m^2 - 2mr - 2r^2 \big)  \\
& \qquad + (p+1)^2 e^{\tfrac{4m}{r}} 
     \big( m^2 \cos(2\theta) - m^2 + 2mr - 2r^2 \big)  \\
& \qquad + 4m (p^2 - 1) r e^{\tfrac{2m}{r}}
   \Big)^2 \\[6pt]
& + 4 \Big( 
     (p-1)^2 \big( m^2 \sin^2(\theta) + r(m+r) \big) \\
& \qquad - (p+1)^2 e^{\tfrac{4m}{r}} \big( m^2 \sin^2(\theta) + r(r-m) \big) \\
& \qquad + 2m (p^2 - 1) r e^{\tfrac{2m}{r}}
   \Big)^2 \\[6pt]
& + 4 \Big( 
     (p-1)^2 \big( m^2 \sin^2(\theta) + 2r(m+r) \big) \\
& \qquad - (p+1)^2 e^{\tfrac{4m}{r}} \big( m^2 \sin^2(\theta) + 2r(r-m) \big) \\
& \qquad + 8m (p^2 - 1) r e^{\tfrac{2m}{r}}
   \Big)^2 \\[6pt]
& + \Big( 
    (p+1)^2 e^{\tfrac{4m}{r}} \big( m^2 \cos(2\theta) - (m-2r)^2 \big) \\
& \qquad - (p-1)^2 \big( m^2 \cos(2\theta) - (m+2r)^2 \big)
   \Big)^2 \\[6pt]
& + 8r^2 \Big(
     m \big( (p+1) e^{\tfrac{2m}{r}} + p - 1 \big)^2 
     + r \big( (p-1)^2 - (p+1)^2 e^{\tfrac{4m}{r}} \big)
   \Big)^2 
\end{aligned}
\end{equation}

Note that when we set $p=1$, the regular polynomial becomes

\begin{equation}
\mathcal{P}(r,\theta)=-m^4+3 m^3 r+m^2 \cos (2 \theta ) \left(m^2-3 m r+3 r^2\right)-9 m^2 r^2+12 m r^3-6 r^4.
\end{equation}

\end{document}